\documentclass[12pt]{article}
\usepackage{putex}
\usepackage{graphicx}

\preprint{PUPT-2265}

\institution{PU}{Joseph Henry Laboratories, Princeton University, Princeton, NJ 08544}

\title{Mimicking the QCD equation of state with a dual black hole}

\authors{Steven S. Gubser and Abhinav Nellore}

\abstract{We present numerical and analytical studies of the equation of state of translationally invariant black hole solutions to five-dimensional gravity coupled to a single scalar.  As an application, we construct a family of black holes that closely mimics the equation of state of quantum chromodynamics at zero chemical potential.}

\date{In affectionate memory of A.~Chamblin \\[20pt] April 2008}

\begin{document}

\maketitle

\tableofcontents

\section{Introduction}
\label{INTRODUCTION}

In the supergravity approximation, the near-extremal D3-brane has equation of state $s \propto T^3$, with a constant of proportionality that is $3/4$ of the free-field value for the dual ${\cal N}=4$ super-Yang-Mills theory \cite{Gubser:1996de}.  The speed of sound is $c_s = 1/\sqrt{3}$, as required by conformal invariance.  On the other hand, the speed of sound of a thermal state in quantum chromodynamics (QCD) has an interesting and phenomenologically important dependence on temperature, with a minimum near the cross-over temperature $T_c$.  Lattice studies of the equation of state are too numerous to cite comprehensively, but they include \cite{Boyd:1996bx} (for pure glue), \cite{Karsch:2001cy} (a review article), and \cite{Aoki:2005vt,Cheng:2007jq} (recent studies with $2+1$ flavors).

We would like to find a five-dimensional gravitational theory that has black hole solutions whose speed of sound as a function of temperature mimics that of QCD.  We will not try to include chemical potentials or to account for chiral symmetry breaking.  We will not try to include asymptotic freedom, but instead will limit our computation to $T \lsim 4 T_c$ and assume conformal behavior in the extreme UV.  We will not even try to give an account of confinement, except insofar as the steep rise in the number of degrees of freedom near the cross-over temperature $T_c$ is recovered in our setup, corresponding to a minimum of $c_s$ near $T_c$.  We will not try to embed our construction in string theory, but instead adjust parameters in a five-dimensional gravitational action to recover approximately the dependence $c_s(T)$ found from the lattice.  That action is
 \eqn{GravityAction}{
  S = {1 \over 2\kappa_5^2} \int d^5 x \, \sqrt{-g}
    \left[ R - {1 \over 2} (\partial\phi)^2 - V(\phi) \right] \,.
 }
We will not include higher derivative corrections, which would arise from $\alpha'$ and loop corrections if the theory \eno{GravityAction} were embedded explicitly in string theory.

The ansatz we will study is
 \eqn{BackgroundAnsatz}{
  ds^2 = e^{2A} (-h dt^2 + d\vec{x}^2) + e^{2B} {dr^2 \over h} \,.
 }
where $A$, $B$, and $h$ are functions of $r$, and $\phi$ is also some function of $r$.  This ansatz is dictated by the symmetries: we want translation invariance in the ${\bf R}^{3,1}$ directions parametrized by $(t,\vec{x})$, and we want $SO(3)$ symmetry in the $\vec{x}$ directions but not $SO(3,1)$ boost invariance---because boost invariance is broken by finite temperature.  Assuming conformal behavior in the extreme UV means that we assume the geometry \eno{BackgroundAnsatz} is asymptotically anti-de Sitter.  A regular horizon arises when $h$ has a simple zero.  Let's say the first such zero (that is, the one closest to the conformal boundary) is at $r=r_H$.  It is assumed that $A$ and $B$ are finite and regular at $r=r_H$.  Standard manipulations lead to the following formulas for entropy density and temperature:
 \eqn{EntropyAndTemperature}{
  s = {2\pi \over \kappa_5^2} e^{3A(r_H)} \qquad
  T = {e^{A(r_H) - B(r_H)} |h'(r_H)| \over 4\pi} \,,
 }
and once these quantities are known, the speed of sound can be read off from
 \eqn{SpeedOfSound}{
  c_s^2 = {d\log T \over d\log s} \,.
 }
The formula for the entropy density in \eno{EntropyAndTemperature} comes from the Bekenstein-Hawking result $S = A/4G_N$, where $A$ is the area of the horizon (really a volume in our case) and $G_N = \kappa_5^2/8\pi$.  The formula for the temperature comes from Hawking's result $T = \kappa/2\pi$ where $\kappa$ is the surface gravity at the horizon.

By adjusting $V(\phi)$ one might expect to be able to recover any pre-specified $c_s(T)$, at least within certain limits---perhaps including that $s(T)/T^3$ should be monotonic or some similar criterion (in this connection see \cite{Appelquist:1999hr}).\footnote{An earlier study \cite{Kajantie:2006hv} of thermodynamic properties of putative holographic duals to QCD starts with a lagrangian including an unspecified matter term.}  The main aim of this paper is to characterize how $V(\phi)$ translates into $c_s(T)$ and vice versa.  In section~\ref{CHAMBLINREALL} we begin with the simplest possible case: $c_s(T)$ constant.  It translates into $V(\phi) = V_0 e^{\gamma\phi}$ for some $V_0<0$ and $\gamma$ related to $c_s$.  In section~\ref{MASTER} we tackle the general case, exploiting a weak form of integrability of the equations resulting from plugging \eno{BackgroundAnsatz} into \eno{GravityAction}.  In section~\ref{EXAMPLES} we exhibit several examples.  These include a particular $V(\phi)$ whose corresponding $c_s(T)$ curve closely mimics that of QCD.  We close with a discussion in section~\ref{DISCUSSION}.

The results in this paper are based in large part on \cite{NelloreAdvanced}, and aspects of them will also be summarized in \cite{GNPRshort}.

\section{Chamblin-Reall solutions and an adiabatic generalization of them}
\label{CHAMBLINREALL}

In a $D$-dimensional conformal field theory (meaning a CFT in $D-1$ spatial dimensions plus one time dimension), the entropy density must obey
 \eqn{sConformal}{
  s \propto T^{D-1} \,,
 }
simply because this expression is dimensionally correct and there is no scale other than the temperature that would permit a more complicated dependence.  So the speed of sound is $c_s = 1/\sqrt{D-1}$.  If $D>4$, then we could obtain a non-conformal theory in four dimensions by compactifying our ${\rm CFT}_D$ on a $D-4$-dimensional torus.  (A similar idea has been considered in \cite{Benincasa:2006ei,Buchel:2007mf}.)  Doing so should not change the speed of sound: a planar sound wave in the resulting $4$-dimensional theory would correspond to a planar sound wave in the original theory whose propagation is in the direction of the uncompactified directions.

The $AdS_{D+1}$-Schwarzschild solution is an extremum of the action
 \eqn{HigherDimAction}{
  S = {1 \over 2\kappa_{D+1}^2} \int d^{D+1} x \, \sqrt{-\hat{g}}
   \left[ \hat{R} + {D(D-1) \over L^2} \right] \,,
 }
and it takes the form
 \eqn{HigherDimAdSSch}{
  d\hat{s}^2 = {L^2 \over z^2} \left( -h dt^2 + d\widehat{\vec{x}}^2 +
    {dz^2 \over h} \right) \,,
 }
where
 \eqn{HigherDimHfct}{
  h = 1 - {z^D \over z_H^D} \,.
 }
We use hats to distinguish $D+1$-dimensional quantities from $4$-dimensional ones.  It is easy to see that $T \propto 1/z_H$ and $s \propto 1/z_H^{D-1}$, so that $s \propto T^{D-1}$ as the conformal field theory requires.  Suppose we now perform the dimensional reduction described in the previous paragraph on the solution \eno{HigherDimAdSSch}.  In slightly more generality than we need, the Kaluza-Klein ansatz is
 \eqn{KKansatz}{
  d\hat{s}^2 = \exp\left\{ \sqrt{{2 \over 3} {D-4 \over D-1}} \, \phi
    \right\} ds^2 +
   \exp\left\{ -\sqrt{6 \over (D-1)(D-4)} \, \phi \right\} ds_{D-4}^2 \,,
 }
where $ds^2$ is a five-dimensional metric and $ds_{D-4}^2$ is the flat metric on a torus ${\bf T}^{D-4}$, whose shape we will assume to be square with side length $\ell$, so that $\Vol {\bf T}^{D-4} = \ell^{D-4}$.  All components of the metric, and also $\phi$, are assumed to depend only on the five-dimensional coordinates.  It is assumed that $\ell$ is a constant; variation of the size of the torus is taken care of by the exponential prefactor multiplying $ds_{D-4}^2$ in \eno{KKansatz}.  The particular coefficients in the exponentials were chosen presciently to obtain a simple five-dimensional action.  Comparing the general form \eno{KKansatz} with the specific solution \eno{HigherDimAction}, one finds
 \eqn{CRsoln}{
  ds^2 = \left( {L \over z} \right)^{{2 \over 3} (D-1)}
    \left( -h dt^2 + d\vec{x}^2 + {dz^2 \over h} \right)
    \qquad e^\phi = \left( {z \over L} \right)^{
    \sqrt{{2 \over 3} (D-1)(D-4)}} \,,
 }
where $h = 1-z^D/z_H^D$ as in \eno{HigherDimHfct}.  The line element \eno{CRsoln} was obtained by the authors of \cite{Chamblin:1999ya}, but not via Kaluza-Klein reduction; instead, they considered black hole solutions to the equations of motion from an action like \eno{GravityAction} with potentials of the form
 \eqn{CRform}{
  V(\phi) = V_0 e^{\gamma\phi} \,,
 }
with $V_0 < 0$.  To see that the solutions have to come out the same in either approach, let's carry through the Kaluza-Klein reduction at the level of the action by plugging \eno{KKansatz} into \eno{HigherDimAction}.  After performing the trivial integral over $T^{D-4}$, one obtains
 \eqn{KKaction}{
  S = {\ell^{D-4} \over 2\kappa_{D+1}^2} \int
    d^5 x \, \sqrt{-g} \left[ R - {1 \over 2} (\partial\phi)^2 -
     V(\phi) \right] \,,
 }
where $V(\phi)$ has the form \eno{CRform} with the identifications
 \eqn{CRidentify}{
  V_0 = -{D(D-1) \over L^2} \qquad
  \gamma = \sqrt{{2 \over 3} {D-4 \over D-1}} \,.
 }
Evidently, the length scale $\ell$ enters the action only as a prefactor, which can be absorbed into a definition of the five-dimensional gravitational constant: $\kappa_5^2 = \kappa_{D+1}^2/\ell^{D-4}$.

By comparing the expression for $\gamma$ in \eno{CRidentify} with the result $c_s = 1/\sqrt{D-1}$ for the speed of sound, we find
 \eqn{csCR}{
  c_s^2 = {1 \over 3} - {\gamma^2 \over 2} \,.
 }
This result can be derived more directly by showing that $s \propto T^{6/(2-3\gamma^2)}$ for Chamblin-Reall solutions: explicitly,
 \eqn{sAndTCR}{
  s &= {1 \over 2\kappa_5^2} \left( {L \over z_H} \right)^{D-1}
    = {1 \over 2\kappa_5^2}
     \exp\left\{-{\phi_H \over \gamma} \right\}
     \cr\noalign{\vskip2\jot}
  T &= {D \over 4\pi z_H} = {1 \over 4\pi L} {8-3\gamma^2 \over
      2-3\gamma^2}
      \exp\left\{ \left( {\gamma \over 2} - {1 \over 3\gamma}
        \right) \phi_H \right\} \,,
 }
where $\phi_H$ is the value of $\phi$ at the horizon.  The dimensional reduction we have described is well-defined only for integer $D>4$, but for the purposes of the computations presented here, it can be any real number greater than $4$.

Suppose we rewrite the result \eno{sAndTCR} as
 \eqn{sAndTCRagain}{
  \log s &= -{\phi_H \over \gamma} + \hbox{(constant in $\phi_H$)}
    \cr
  \log T &= \left( {\gamma \over 2} - {1 \over 3\gamma} \right)
    \phi_H + \hbox{(constant in $\phi_H$)} \,.
 }
Given \eno{sAndTCRagain} and the formula $\gamma = V'(\phi)/V(\phi)$, a natural next step would be to guess the following dependence of $s$ and $T$ on $\phi_H$ when $\gamma$ is a slowly varying function of $\phi$ rather than a constant:
 \eqn{sAndTguess}{
  \log s &= -\int_{\phi_0}^{\phi_H} d\phi \,
   {V(\phi) \over V'(\phi)} +
    \hbox{(slowly varying in $\phi_H$)}  \cr
  \log T &= \int_{\phi_0}^{\phi_H} d\phi \, \left(
    {1 \over 2} {V'(\phi) \over V(\phi)} -
    {1 \over 3} {V(\phi) \over V'(\phi)} \right) +
    \hbox{(slowly varying in $\phi_H$)} \,.
 }
The lower limit $\phi_0$ in the integrals is an arbitrary cutoff.  If we assume that $V(\phi)$ has a maximum at $\phi=0$ and an expansion of the form \eno{Vexpand}, then $V(\phi)/V'(\phi) \approx -12/(m^2 L^2 \phi)$ near $\phi=0$.  So the integrals in \eno{sAndTguess} diverge if they are continued all the way to $\phi=0$, and the cutoff $\phi_0$ must be chosen to have the same sign as $\phi_H$ to avoid this divergence.

A consequence of the estimates \eno{sAndTguess} is a simple formula for the speed of sound:
 \eqn{csEstimate}{
  c_s^2 = {d\log T / d\phi_H \over d\log s / d\phi_H}
    \approx {1 \over 3} - {1 \over 2}
     {V'(\phi_H)^2 \over V(\phi_H)^2} \,.
 }
Another consequence is
 \eqn{DOFguess}{
  \log {s \over T^3} &= -{3 \over 2} \int_{\phi_0}^{\phi_H} d\phi \,
    {V'(\phi) \over V(\phi)} +
    \hbox{(slowly varying in $\phi_H$)}  \cr
   &= -{3 \over 2} \log {V(\phi_H) \over V(\phi_0)} +
    \hbox{(slowly varying in $\phi_H$)} \,.
 }
A simpler way of expressing \eno{DOFguess} is
 \eqn{DOFguessFinal}{
  {s \over T^3} \propto |V(\phi_H)|^{-3/2} \,,
 }
up to corrections from slowly varying terms.  This is interesting because $s/T^3$ is one way of defining the effective number of degrees of freedom available to a system, and we see from \eno{DOFguessFinal} that it is closely related to the potential evaluated at the horizon.

The results \eno{csEstimate} and~\eno{DOFguess} are a first attempt at solving the problem of translating an arbitrary $V(\phi)$ to an equation of state, or an arbitrary equation of state into $V(\phi)$.  Here's how the latter process would work.  Suppose one specifies the equation of state as $s=s(T)$.  Ignoring corrections to \eno{DOFguess}, one has
 \eqn{fDef}{
  f \equiv -{2 \over 3} \log {s \over T^3} =
    \log {V \over V_0} \,,
 }
where $V_0$ is some constant.  Let's regard $f$ as the independent variable.  Because $V = V_0 e^f$, all we need is to find $\phi=\phi(f)$, and we will have a parametric representation of $V(\phi)$.  One may rewrite \eno{csEstimate} as
 \eqn{fUse}{
  c_s^2 = {1 \over 3} - {1 \over 2 (d\phi/df)^2} \,,
 }
where corrections have again been ignored.  Knowing $s(T)$ with good precision means one can express $c_s^2$ as a function of $f$.  Then \eno{fUse} can readily be integrated to give
 \eqn{phiInt}{
  \phi(f) = \int {df \over \sqrt{2 \left(
    {1 \over 3} - c_s(f)^2 \right)}} \,.
 }
The integral is left in indefinite form because adding a constant to $\phi$ is obviously allowed.

We stress that the result of plugging \eno{phiInt} into the form $V=V_0 e^f$ will result in a $V(\phi)$ that only approximately reproduces the desired $s(T)$.  If the speed of sound varies rapidly with $T$, the approximation may be poor.  In section~\ref{APPROXIMATE} we will show how to improve this approximation without resorting to differential equations that cannot be explicitly solved in terms of indefinite integrals.

\section{A nonlinear master equation}
\label{MASTER}

There is a residual gauge freedom in the ansatz \eno{BackgroundAnsatz}, namely reparametrization of the radial direction.  A convenient gauge choice, which should be at least piecewise valid in any geometry where the scalar is non-vanishing, is to set $r=\phi$.  Then the line element becomes
 \eqn{phiGauge}{
  ds^2 = e^{2A} (-hdt^2 + d\vec{x}^2) + e^{2B} {d\phi^2 \over h} \,,
 }
and the equations of motion following from the action \eno{GravityAction} take the form
 \begin{subequations}\label{BackgroundEoms}\begin{align}
 \label{BackgroundEom1}
  A'' - A' B' + {1 \over 6} &= 0  \\
 \label{BackgroundEom2}
  h'' + (4A'-B') h' &= 0  \\
 \label{BackgroundEom3}
  6 A' h' + h (24A'^2-1) + 2 e^{2B} V &= 0  \\
 \label{BackgroundEom4}
  4A' - B' + {h' \over h} - {e^{2B} \over h} V' &= 0 \,,
 \end{align}\end{subequations}
where primes denote $d/d\phi$.  The first two of these equations come from the $tt$ and $x^1x^1$ Einstein equations; the third comes from the $\phi\phi$ Einstein equation; and the last comes from the scalar equation of motion.  There is typically some redundancy in equations obtained from classical gravity, with or without matter.  In the case of \eno{BackgroundEoms}, the redundancy is that the $\Phi$ derivative of the third equation follows algebraically from the four equations listed.

The ansatz \eno{phiGauge} has one peculiar feature: $e^{2B}$ must have dimensions of length squared.  This is because $\phi$ is dimensionless.

The equations of motion \eno{BackgroundEoms} enjoy a weak form of integrability, in the following sense: If a smooth ``generating function'' $G(\phi)$ is specified, then it is possible to find a black hole solution where $A'(\phi) = G(\phi)$ in terms of indefinite integrals of simple functions of $G(\phi)$ and $G'(\phi)$.  But $V(\phi)$ itself is expressed in terms of such integrals, and one cannot easily find all the possible $G(\phi)$ that lead to a specified $V(\phi)$.  In other words, there can be simple analytic solutions to \eno{BackgroundEoms} for special $V(\phi)$ at a special value of the temperature, but as far as we know, there is no nontrivial $V(\phi)$ (i.e., none besides the exponential form) for which analytic solutions exist over a continuous range of temperatures.

To understand this claim of integrability, let us consider $A'(\phi)=G(\phi)$ to be fixed as a function of $\phi$ and work out $A(\phi)$, $B(\phi)$, $h(\phi)$, and $V(\phi)$.  The first of these is trivial:
 \eqn{TrivialA}{
  A(\phi) = A_0 + \int_{\phi_0}^\phi d\tilde\phi \, G(\tilde\phi) \,.
 }
Computing $B(\phi)$ is immediate once one solves \eno{BackgroundEom1} for $B'$:
 \eqn{TrivialB}{
  B(\phi) = B_0 + \int_{\phi_0}^\phi d\tilde\phi \,
    {G'(\tilde\phi) + 1/6 \over G(\tilde\phi)} \,.
 }
Next one observes that \eno{BackgroundEom2} is straightforwardly solved once one knows $A(\phi)$ and $B(\phi)$:
 \eqn{TrivialH}{
  h(\phi) = h_0 + h_1 \int_{\phi_0}^\phi d\tilde\phi \,
    e^{-4A(\tilde\phi) + B(\tilde\phi)} \,.
 }
Now \eno{BackgroundEom3} can be solved for $V(\phi)$ in terms of known quantities:
 \eqn{TrivialV}{
  V(\phi) = h(\phi) {e^{-2B(\phi)} \over 2} \left( 1 -
    24 G(\phi)^2 - 6 G(\phi) {h'(\phi) \over h(\phi)} \right) \,.
 }
The constraint equation \eno{BackgroundEom4} doesn't yield any new information.

If one chooses
 \eqn{Gphi}{
  G(\phi) = -{1 \over 3\gamma} \,,
 }
then by working through \eno{TrivialA}-\eno{TrivialV} one recovers the Chamblin-Reall solution in the form
 \eqn{CRphi}{
  A(\phi) &= A_0 - {\phi-\phi_0 \over 3\gamma}  \cr
  B(\phi) &= B_0 - {\gamma \over 2} (\phi-\phi_0) \cr
  h(\phi) &= h_0 + \tilde{h}_1 \exp\left\{
   {8-3\gamma^2 \over 6\gamma} (\phi-\phi_0) \right\}  \cr
  V(\phi) &= V_0 e^{\gamma\phi}
 }
where
 \eqn{NastyConstants}{
  \tilde{h}_1 = {6 e^{-4A_0+B_0} \gamma \over 8-3\gamma^2} h_1 \qquad
  V_0 = -{8-3\gamma^2 \over 6\gamma} e^{-2B_0-\gamma\phi_0} h_0 \,.
 }
By choosing
 \eqn{phiZeroChoice}{
  \phi_0 = {1 \over \gamma} \left( \log h_0 - 2B_0 \right) \,,
 }
one obtains $V(\phi)$ in a form that doesn't depend on any integration constants at all.  This situation is very special and corresponds to the fact that for $V(\phi) \propto -e^{\gamma\phi}$ one can find a whole family of analytic solutions parametrized by $\phi_H$.

By differentiating combinations of \eno{TrivialA}-\eno{TrivialV}, one can derive the following ``nonlinear master equation:''
 \eqn{mastereq}{
  {G' \over G + V/3V'} = {d \over d\phi} \log\left(
    {G' \over G} + {1 \over 6G} - 4G -
    {G' \over G + V/3V'} \right) \,.
 }
Describing \eno{mastereq} as the master equation is appropriate because if one starts knowing $V(\phi)$ and manages to solve \eno{mastereq}, then to obtain a black hole solution one only needs to perform the ``trivial'' integrations \eno{TrivialA}-\eno{TrivialH}.  A numerically efficient strategy for obtaining an equation of state given $V(\phi)$ centers around solving \eno{mastereq} numerically.  In more detail, the procedure is:
 \begin{enumerate}
  \item Choose the value $\phi_H$ of the scalar at the horizon.\label{ChoosePhiH}
  \item Find a series solution of the nonlinear master equation around $\phi=\phi_H$.\label{SeriesSolution}
  \item Seed a numerical integrator like Mathematica's {\tt NDSolve} close to $\phi=\phi_H$ using the series solution.\label{SeedNumerics}
  \item Integrate the nonlinear master equation up to a value of $\phi$ close to a maximum, corresponding to the asymptotically anti-de Sitter part of the geometry.\label{IntegrateMaster}
  \item Extract $s$ and $T$ from integrals of simple functions of $G(\phi)$.\label{ExtractThermo}
 \end{enumerate}
Most of these steps require further explanation, which will occupy the remainder of this section.

At the horizon, $h$ has a simple zero, and the other quantities appearing in \eno{BackgroundEom3} and \eno{BackgroundEom4} are finite.  Evaluating these two equations at the horizon gives
 \eqn{TwoAtHorizon}{
  V(\phi_H) = -3e^{-2B(\phi_H)} G(\phi_H) h'(\phi_H) \qquad
  V'(\phi_H) = e^{-2B(\phi_H)} h'(\phi_H) \,,
 }
which implies that $G + V/3V'$ vanishes at the horizon.  Starting from this condition, one may develop a power series solution around the horizon:
 \eqn{horizonexpand}{
  G(\phi) = -{1 \over 3}{V(\phi_{H}) \over V'(\phi_{H})} +
    {1 \over 6} \left({V(\phi_{H})V''(\phi_{H}) \over
    V'(\phi_{H})^{2}}-1\right)
    (\phi-\phi_{H}) + O\left[ (\phi-\phi_H)^2 \right] \,.
 }
This series solution can be developed to any desired order without encountering arbitrary integration constants.

To understand the asymptotic behavior far from the horizon, let's specialize to the case where $V(\phi)$ has a maximum at $\phi=0$:
 \eqn{Vexpand}{
  V(\phi) = -{12 \over L^2} + {1 \over 2} m^2 \phi^2 +
   O(\phi^3) \,,
 }
where $m^2 < 0$ in order for $\phi=0$ to be a maximum.  The gauge theory operator ${\cal O}_\phi$ dual to $\phi$ has dimension $\Delta$, where
 \eqn{DeltaFromM}{
  \Delta (\Delta-4) = m^2 L^2 \,.
 }
We will be primarily interested in the case where $\Delta$ is close to $4$.  It helps our intuition at this point to pass to a more standard gauge: instead of setting $r=\phi$, we can set $B=0$ to obtain
 \eqn{StandardGauge}{
  ds^2 = e^{2A} (-hdt^2 + d\vec{x}^2) + {dr^2 \over h} \,.
 }
We note that $A$ and $h$ appearing in \eno{StandardGauge} are precisely the same as when we use the $r=\phi$ gauge, only expressed as functions of $r$ rather than $\phi$.  Large $r$ corresponds to the region far from the horizon, and the leading asymptotic behavior of solutions there is
 \eqn{FarAsymptotics}{
  A \approx {r \over L} \qquad
  h \approx 1 \qquad
  \phi \approx (\Lambda L)^{4-\Delta} e^{(\Delta-4)A}\,.
 }
Each approximate equality in \eno{FarAsymptotics} means that the ratio of the two expressions on each side approaches $1$ as $r$ becomes large.  The behavior of $\phi$ indicates a relevant deformation of the conformal field theory:
 \eqn{DeformCFT}{
  {\cal L} = {\cal L}_{\rm CFT} + \Lambda^{4-\Delta} {\cal O}_\phi \,.
 }
As we vary temperature to compute the equation of state, we should of course keep $\Lambda$ fixed.  A simple way to do this is to set $\Lambda L=1$.  Then the last equation in \eno{FarAsymptotics} becomes
 \eqn{aAsymp}{
  A(\phi) = {\log\phi \over \Delta-4} + o(1) \qquad
   \hbox{for small $\phi$,}
 }
where $o(1)$ means a quantity that is parametrically smaller than $1$ in the limit under consideration---so in the limit $\phi \to 0$ in the case of \eno{aAsymp}.

In order to compute the entropy density using \eno{EntropyAndTemperature}, we need to know $A(\phi_H)$.  This can be extracted by comparing \eno{aAsymp} to \eno{TrivialA} with $\phi_0$ set equal to $\phi_H$ and $A_0$ set equal to $A_H = A(\phi_H)$:
 \eqn[c]{TwoForms}{
  A(\phi) = A_H + \int_{\phi_H}^\phi d\tilde\phi \, G(\tilde\phi)
    = {\log\phi \over \Delta-4} + o(1) \,.
 }
Solving for $A_H$ and then taking $\phi \to 0$, one finds
 \eqn{FoundAH}{
  A_H = {\log\phi_H \over \Delta-4} +
    \int_0^{\phi_H} d\phi \left[ G(\phi) - {1 \over
     (\Delta-4)\phi} \right] \,.
 }
The integral converges because the explicit $1/\phi$ term cancels against the leading behavior of $G(\phi)$ for small $\phi$.  Plugging \eno{FoundAH} into the expression for entropy density from \eno{EntropyAndTemperature}, we find at last
 \eqn{FoundEntropy}{
  s = {2\pi \over \kappa_5^2} \phi_H^{3/(\Delta-4)}
    \exp\left\{ 3 \int_0^{\phi_H} d\phi \,
      \left[ G(\phi) - {1 \over (\Delta-4)\phi} \right]
      \right\} \,.
 }

A similar formula for the temperature may be derived starting with the observation that
 \eqn{drdphi}{
  {dr \over d\phi} = -e^B \,,
 }
where $B$ is the function controlling the $g_{\phi\phi}$ metric component in $r=\phi$ gauge.  One obtains \eno{drdphi} by comparing the \eno{phiGauge} to \eno{StandardGauge}.  The sign is based on assuming that $\phi$ increases from $0$ to positive values as $r$ decreases from $+\infty$ to finite values.  The first equation in \eno{FarAsymptotics} implies $dA/dr \to 1/L$ as $r \to \infty$.  Combining this with \eno{drdphi} gives
 \eqn{SmallPhiHandle}{
  G = {dA \over d\phi} = {dr \over d\phi} {dA \over dr}
    \approx -e^B {1 \over L} \,,
 }
where the approximate equality means that the ratio of the last expression to the previous ones approaches $1$ as $\phi$ goes to $0$.  In summary,
 \eqn{GBlimit}{
  1 \approx -L G(\phi) e^{-B(\phi)} \,,
 }
using the same sense of approximate equalities.  (Recall that $e^{-B}$ has dimensions of inverse length, while $G(\phi)$ is dimensionless, so \eno{GBlimit} is dimensionally correct.)  We may cast the expression for temperature from \eno{EntropyAndTemperature} in terms of $G(\phi)$ by multiplying by $1$ in the form indicated in \eno{GBlimit}:
 \eqn{CalculateTemperature}{
  T &= {e^{A_H - B(\phi_H)} |h'(\phi_H)| \over 4\pi}
    \approx {e^{A_H - B(\phi_H)} h'(\phi_H) \over 4\pi}
      L G(\phi) e^{-B(\phi)}  \cr
   &= {L e^{-2B(\phi_H)} G(\phi_H) h'(\phi_H) \over 4\pi}
    \exp\left\{ A_H + B(\phi_H) - B(\phi) - \log {G(\phi_H) \over
      G(\phi)} \right\}  \cr
   &= -{LV(\phi_H) \over 12\pi}
     \exp\left\{ A_H + \int_\phi^{\phi_H} {d\tilde\phi \over
       6G(\tilde\phi)} \right\} \,.
 }
In the second step we assumed that $h'(\phi_H)<0$, which is the expected sign when $\phi$ vanishes on the boundary and is positive at the horizon.  In the last step we used the first equation from \eno{TwoAtHorizon} to simplify the prefactor and also
 \eqn{Bdiff}{
  B(\phi_H) - B(\phi) = \log {G(\phi_H) \over G(\phi)} +
    \int_\phi^{\phi_H} {d\tilde\phi \over 6G(\tilde\phi)} \,,
 }
which is a consequence of \eno{TrivialB}.  The integral in the last expression of \eno{FoundTemperature} converges even when $\phi \to 0$.  We use \eno{FoundAH} to eliminate $A_H$ from \eno{FoundTemperature} and obtain at last
 \eqn{FoundTemperature}{
  T = {\phi_H^{1/(\Delta-4)} \over \pi L} {V(\phi_H) \over V(0)}
    \exp\left\{ \int_0^{\phi_H} d\phi\left[ G(\phi) -
      {1 \over (\Delta-4)\phi} + {1 \over
        6G(\phi)} \right] \right\} \,.
 }
The measure of the number of degrees of freedom that is easiest for us to access is
 \eqn{soverTcubed}{
  {s \over T^3} = 2\pi^4 {L^3 \over \kappa_5^2}
    {V(0)^3 \over V(\phi_H)^3}
    \exp\left\{ -3 \int_0^{\phi_H} {d\phi \over 6G(\phi)}
      \right\} \,,
 }
where to obtain the right hand side we simply combined \eno{FoundEntropy} and \eno{FoundTemperature}.  When $\phi_H$ is small, entropy and temperature become large because of the factors $\phi_H^{3/(\Delta-4)}$ and $\phi_H^{1/(\Delta-4)}$ in \eno{FoundEntropy} and \eno{FoundTemperature}.  In this limit, the integrals in the exponent become negligible, and \eno{soverTcubed} becomes
 \eqn{HighTLimitS}{
  {s \over T^3} \approx 2\pi^4 {L^3 \over \kappa_5^2} \,.
 }
So we recover conformal behavior in the ultraviolet, as expected.

\section{An approximate determination of the equation of state}
\label{APPROXIMATE}

\begin{figure}
  \centerline{\includegraphics[width=7in]{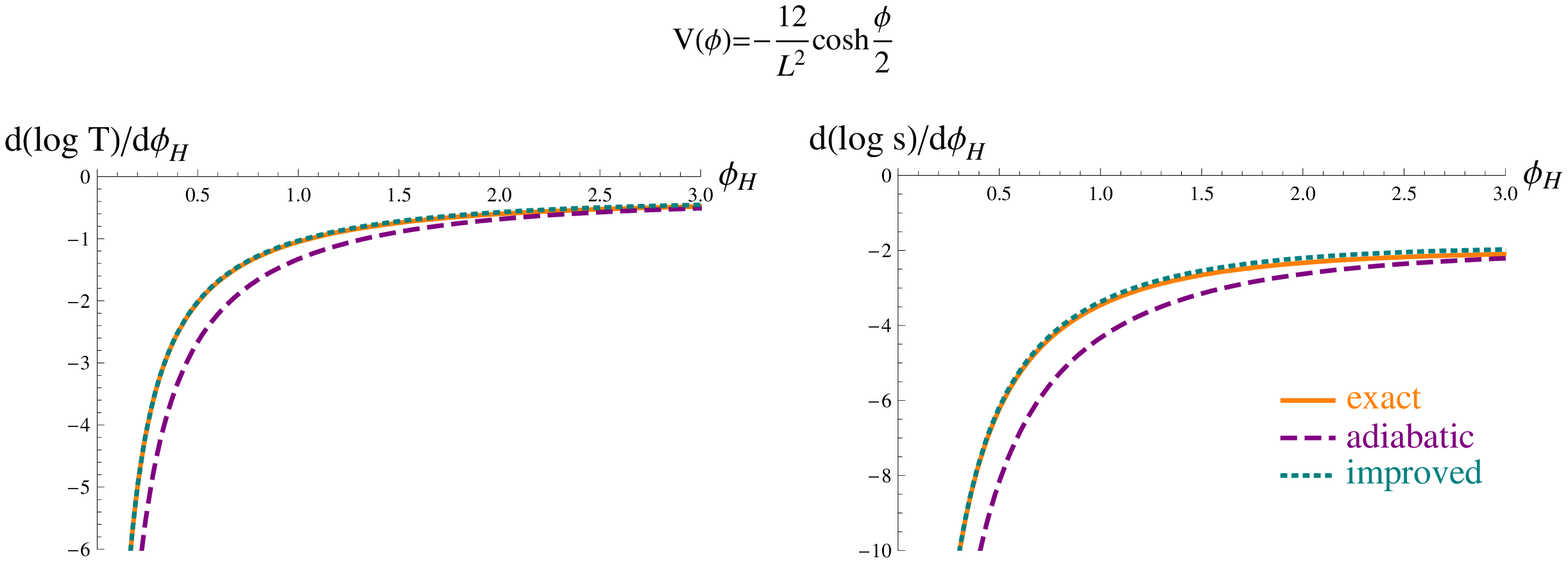}}
  \caption{A comparison of the exact ${d(\log s) \over d\phi_H}$ and ${d(\log T) \over d\phi_H}$ for $V(\phi)=-{12\over L^2}\cosh{\phi\over 2}$ with the adiabatic approximation, \eno{sAndTguess}, and the improved approximation scheme, \eno{Fudgy} with the choice \eno{GuessedInterpolator}.}\label{TwoApprox}
 \end{figure}

The adiabatic formulas \eno{sAndTguess} work well when $\phi_H$ is in a region where $V(\phi)$ is nearly exponential, but they do not work well for small $\phi_H$, where $V(\phi)$ is close to attaining a maximum.  This is shown in figure~\ref{TwoApprox} for $V(\phi)=-(12/L^2)\cosh(\phi/2)$.
On the other hand, it's easy to extract asymptotic formulas valid in the $\phi_H \to 0$ limit from~\eno{FoundEntropy} and~\eno{FoundTemperature}: using the expansion \eno{Vexpand}, one finds
 \eqn{UVsandT}{
  T = {1 \over \pi L}\phi_H^{1/(\Delta-4)} \qquad
  s ={2\pi \over \kappa_5^2} \phi_H^{3/(\Delta-4)} \,.
 }
We wish to find formulas that interpolate smoothly between \eno{sAndTguess} and~\eno{UVsandT} and involve at most indefinite integrals, not solutions to a difficult, nonlinear, second-order differential equation such as \eno{mastereq}.  We start by noting that the formulas \eno{UVsandT} together with \eno{Vexpand} imply
 \eqn{approxSmallPhiH}{
  {d\log T \over d\phi_H} &\approx {\Delta \over 4}
    \left( {1 \over 2} {V'(\phi_H) \over V(\phi_H)} -
      {1 \over 3} {V(\phi_H) \over V'(\phi_H)} \right)  \cr
  {d\log s \over d\phi_H} &\approx {\Delta \over 4}
    \left( -{V(\phi_H) \over V'(\phi_H)} \right) \qquad
    \hbox{for small $\phi_H$,}
 }
where approximate equality means that the ratio of the two sides tends to $1$ as $\phi_H \to 0$.  On the other hand, provided $V(\phi)$ tends to an exponential form $V_0 e^{\gamma\phi}$ for large $\phi$, the adiabatic approximation becomes good if $\phi_H$ is large.  So for such potentials, \eno{sAndTguess} can be rephrased as
 \eqn{approxLargePhiH}{
  {d\log T \over d\phi_H} &\approx
    {1 \over 2} {V'(\phi_H) \over V(\phi_H)} -
      {1 \over 3} {V(\phi_H) \over V'(\phi_H)}  \cr
  {d\log s \over d\phi_H} &\approx
    -{V(\phi_H) \over V'(\phi_H)} \qquad
    \hbox{for large $\phi_H$.}
 }
Comparing \eno{approxSmallPhiH} and~\eno{approxLargePhiH}, we are led to introduce ``fudge factors'' $\rho_s(\phi_H)$ and $\rho_T(\phi_H)$ such that
 \eqn{Fudgy}{
  {d\log T \over d\phi_H} &= \rho_T(\phi_H)
    \left( {1 \over 2} {V'(\phi_H) \over V(\phi_H)} -
      {1 \over 3} {V(\phi_H) \over V'(\phi_H)} \right)  \cr
  {d\log s \over d\phi_H} &= \rho_s(\phi_H)
    \left( -{V(\phi_H) \over V'(\phi_H)} \right) \,.
 }
We can rephrase \eno{approxSmallPhiH} and~\eno{approxLargePhiH} as the statements that both $\rho_T(\phi_H)$ and $\rho_s(\phi_H)$ interpolate between $\Delta/4$ at small $\phi_H$ and $1$ at large $\phi_H$.  Our improved estimate of the equation of state consists simply of guessing an interpolating function with these properties.  The guess is
 \eqn{GuessedInterpolator}{
  \rho_T(\phi_H) \approx \rho_s(\phi_H) \approx \rho(\phi_H) \equiv
    1 + {V(0) \over V(\phi_H)} \left( {\Delta \over 4} - 1 \right)
    \,.
 }
In \eno{GuessedInterpolator} approximate equality means that $\rho_T$, $\rho_s$, and $\rho$ are supposed to be nearly equal for all $\phi_H$.  But away from the small $\phi_H$ and large $\phi_H$ limits, \eno{GuessedInterpolator} is not a controlled approximation, in the sense that there isn't a parameter that we can tune to make it better.  It is nevertheless useful for quickly determining the qualitative features of an equation of state given $V(\phi)$, as illustrated in figure~\ref{TwoApprox}.  There might be a better choice of $\rho_T(\phi_H)$ and $\rho_s(\phi_H)$, or perhaps even a systematic expansion for them in terms of powers of the potential and its derivatives.

Starting from \eno{Fudgy}, we have immediately
 \eqn{csSame}{
  c_s^2 = {d\log T / d\phi_H \over d\log s / d\phi_H} =
    {\rho_T(\phi_H) \over \rho_s(\phi_H)}
    \left( {1 \over 3} - {1 \over 2} {V'(\phi_H)^2 \over
      V(\phi_H)^2} \right) \,.
 }
Thus, assuming $\rho_T(\phi_H) \approx \rho_s(\phi_H)$ is the same as assuming that the speed of sound, as a function of $\phi_H$, is well-approximated by the adiabatic formula, \eno{csEstimate}.

\section{Examples}
\label{EXAMPLES}

\begin{figure}
  \centerline{\includegraphics[width=4in]{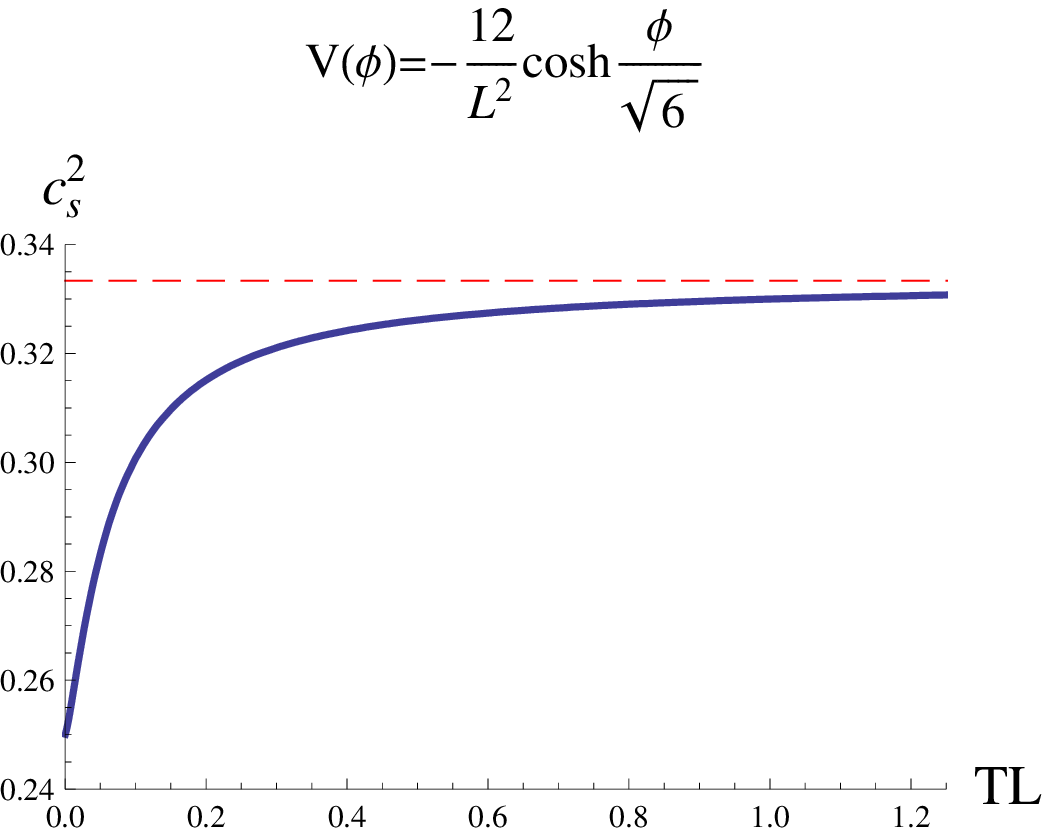}}
  \caption{The speed of sound for $V(\phi) = -{12 \over L^2} \cosh {\phi \over \sqrt{6}}$.}\label{ex1}
 \end{figure}

The simplest analytical form that interpolates between a maximum at $\phi=0$ and exponential behavior for large $\phi$ is
 \eqn{Vcosh}{
  V(\phi) = V_{\cosh}(\phi) \equiv -{12 \over L^2} \cosh\gamma\phi \,.
 }
The adiabatic treatment discussed in section~\ref{CHAMBLINREALL} leads us to expect that the speed of sound will be $c_s = \sqrt{{1\over 3} - {\gamma^2 \over 2}}$ for large $\phi_H$.  So in order to have stable black holes in this regime, we must have $\gamma \leq \sqrt{2/3}$.  This bound can be regarded as an application of the correlated stability conjecture (CSC) \cite{Gubser:2000ec,Gubser:2000mm}.  But there is a tighter bound on $\gamma$ coming from the behavior near $\phi=0$:
 \eqn{VcoshExpand}{
  V_{\cosh}(\phi) = -{12 \over L^2} - {6\gamma^2 \over L^2} \phi^2 +
   O(\phi^4) \,,
 }
so $m^2 = -12\gamma^2/L^2$.  In order to satisfy the Breitenlohner-Freedman bound, $m^2 L^2 \geq -4$ \cite{Breitenlohner:1982bm,Breitenlohner:1982jf,Mezincescu:1984ev}, we must have $\gamma \leq 1/\sqrt{3}$.  This is very restrictive, because it means that the minimum speed of sound we can arrange at large $\phi_H$ using the pure $\cosh$ potential \eno{Vcosh} is $c_s = \sqrt{5/18} \approx 0.53$, only slightly smaller than the conformal value $c_s = 1/\sqrt{3} \approx 0.58$.  The behavior of $c_s^2$ as a function of $T$ is shown in figure~\ref{ex1} for $\gamma = 1/\sqrt{6}$.

\begin{figure}
  \centerline{\includegraphics[width=3.5in]{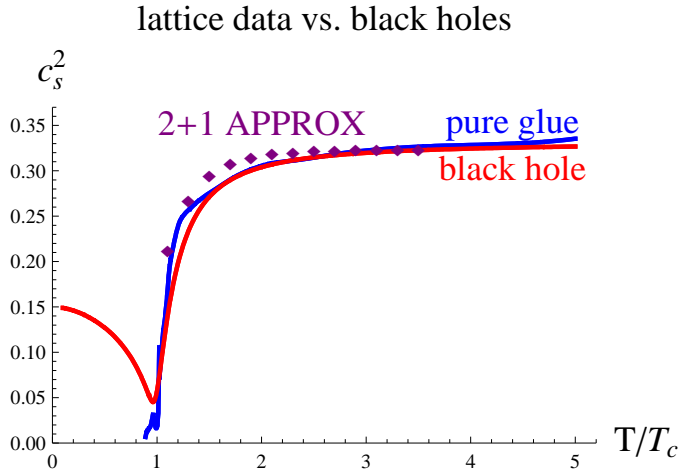}}
  \caption{The equation of state of a black hole (red) compared to the lattice equation of state for pure glue (blue) and $2+1$ QCD.  The pure glue curve is based on \cite{Boyd:1996bx} and private communications from F.~Karsch.  The $2+1$ QCD points are based on \cite{Cheng:2007jq}.}\label{LatticeVsBH}
 \end{figure}

\begin{figure}
  \centerline{\includegraphics[width=7in]{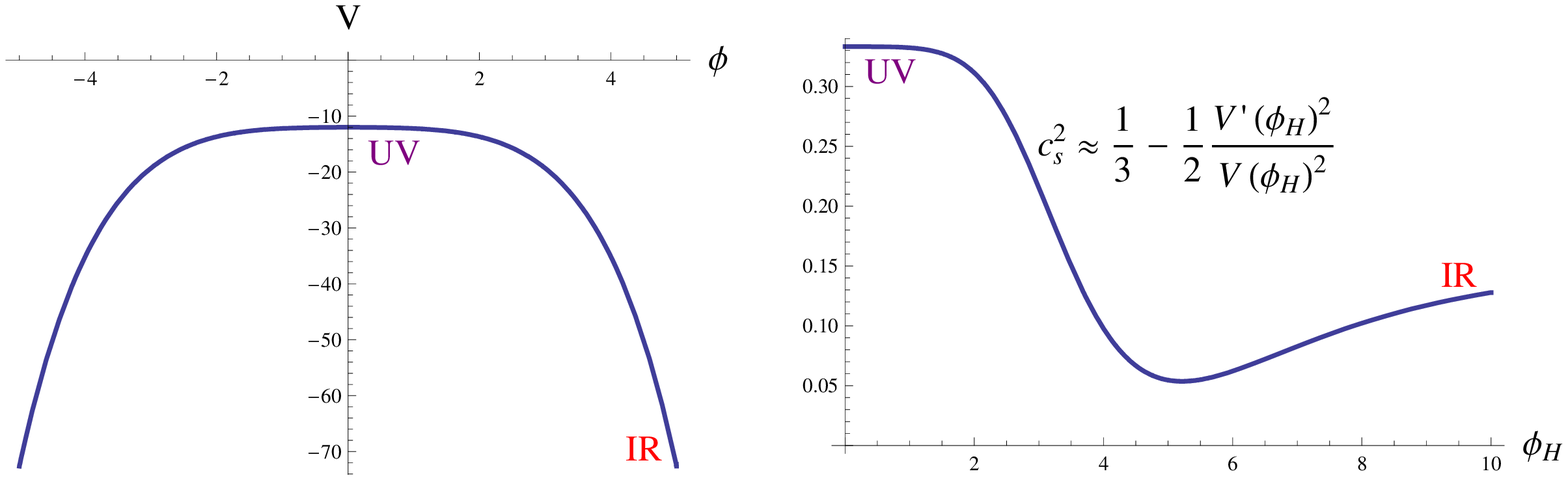}}
  \caption{Left: The potential \eno{bPotential} with the parameter choices \eno{bFromDelta} that give an equation of state resembling QCD's.  Right: Although $V(\phi)$ is relatively featureless, the adiabatic formula $c_s^2 \approx {1 \over 3} - {1 \over 2} {V'(\phi_H)^2 \over V(\phi_H)^2}$ suggests that the equation of state resulting from it will indeed exhibit a low minimum for the speed of sound.}\label{Vfeature}
 \end{figure}

If one uses the potential \eno{Vcosh}, then $c_s$ in the infrared is tied to the dimension $\Delta$ of the operator that breaks conformal invariance in the ultraviolet.  Let us consider a minimal generalization that loosens this artificial constraint:
 \eqn{bPotential}{
  V(\phi) = V(\gamma,b;\phi) \equiv -{12 \over L^2} \cosh\gamma\phi +
    b\phi^2 \,.
 }
A parameter equivalent to $\gamma$, as before, is the speed of sound in the infrared, $c_s^2 = {1 \over 3} - {\gamma^2 \over 2}$.  With $\gamma$ fixed, a parameter equivalent to $b$ is the dimension $\Delta$ of the operator dual to $\phi$ near the UV fixed point:
 \eqn{bFromDelta}{
  b = {6\gamma^2 \over L^2} + {\Delta(\Delta-4) \over 2L^2} \,.
 }
Note that taking $\Delta$ close to $4$ amounts to making $V(\phi)$ nearly quartic near its maximum.  As we will report in more detail in \cite{GNPRshort}, the choice
 \eqn{gammaDeltaChoice}{
  \gamma = 0.606 \qquad b = {2.06 \over L^2} \,,
 }
corresponding to $c_s^2 = 0.15$ in the infrared and $\Delta = 3.93$, leads to an equation of state that bears a striking resemblance to the one expected for QCD: see figure~\ref{LatticeVsBH}.
It may seem surprising that the most distinctive feature of the equation of state of QCD, namely a smooth but rapid cross-over, emerges from a potential that is nearly featureless.  To gain some intuition about why this happened, consider again the adiabatic approximation \eno{csEstimate} to the speed of sound.  When $\phi_H$ is close to where the nearly quartic behavior of $V(\phi)$ rolls over into nearly exponential behavior, this approximate formula predicts that $c_s^2$ dips to a fairly low value, only to rise back up again for larger $\phi_H$ towards its infrared limit, $0.15$.  See figure~\ref{Vfeature}.

\begin{figure}
  \centerline{\includegraphics[width=4in]{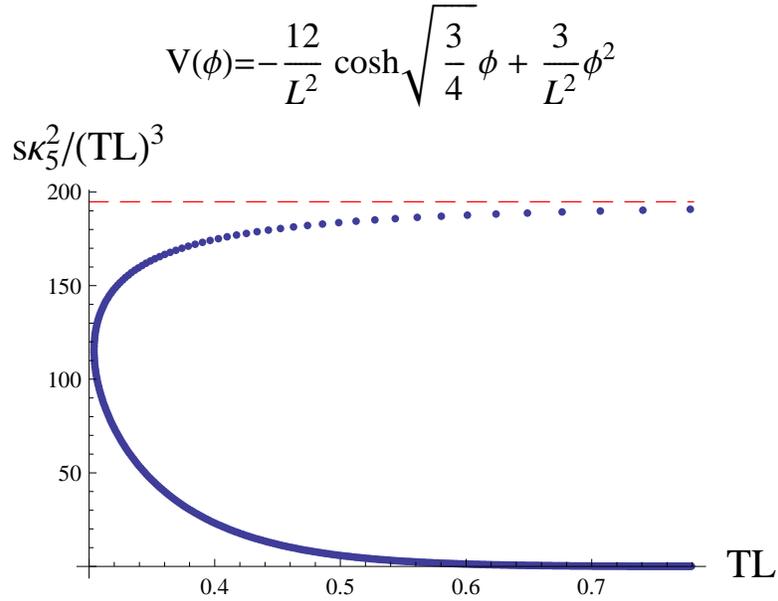}}
  \caption{The equation of state for $V(\phi)=-{12 \over L^2} \cosh \sqrt{3 \over 4} \phi + {3 \over L^2} \phi^2$.}\label{ex2}
 \end{figure}

Other behaviors emerge from the potential \eno{bPotential} for other choices of $b$ and $\gamma$.  For example, if $\gamma > \sqrt{2/3}$, the adiabatic approximation suggests that there is a minimum temperature $T_{\rm min}$ for black hole solutions.  A particular case is illustrated in figure~\ref{ex2}.
Solutions with very low entropy have high temperature and negative specific heat, and they are always thermodynamically disfavored compared to a branch of high-entropy solutions.  Presumably there is a first order transition to geometries with no horizon, similar to the Hawking-Page transition \cite{Hawking:1982dh}.  This transition probably happens at a temperature above $T_{\rm min}$.  It is worth noting that the specific heat diverges at $T_{\rm min}$ because $T$ reaches a minimum as a function of $\phi_H$ while $S$ is varying smoothly.

\begin{figure}
  \centerline{\includegraphics[width=4in]{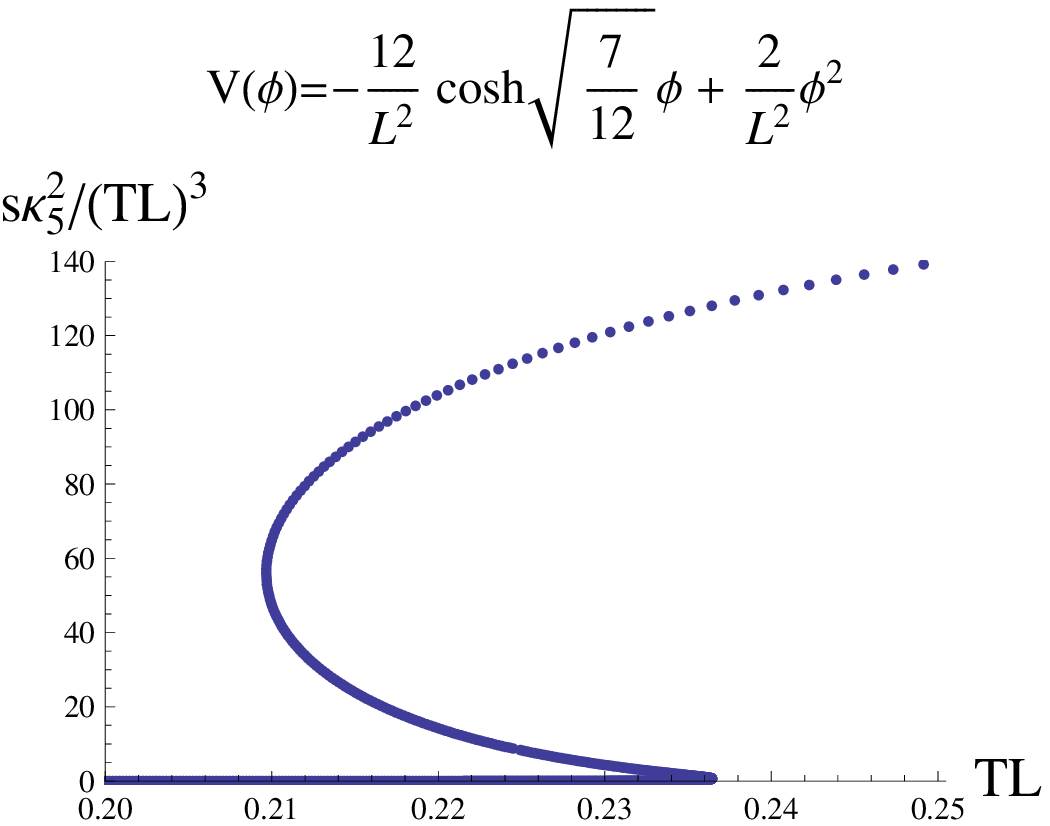}}
  \caption{The equation of state for $V(\phi)=-{12 \over L^2} \cosh \sqrt{7 \over 12} \phi + {2 \over L^2} \phi^2$.}\label{ex3}
 \end{figure}

It is also possible to have a first order transition between high entropy and low entropy black holes.  An example where this happens is illustrated in figure~\ref{ex3}.
For a finite range of $\phi_H$, the speed of sound is imaginary, indicating a Gregory-Laflamme instability.  This touches once again on the correlated stability conjecture (CSC), so let us pause to review it.  It was proposed in \cite{Gubser:2000ec,Gubser:2000mm} and further argued in \cite{Reall:2001ag} that, in the absence of conserved charges related to gauge symmetries, existence of a Gregory-Laflamme instability \cite{Gregory:1993vy,Gregory:1994bj} is equivalent to positivity of the specific heat, $C = T \partial S / \partial T$.  According to a more general version of the CSC, dynamical stability of a horizon is equivalent to positivity of an appropriate Hessian matrix of susceptibilities, one of which is the specific heat \cite{Gubser:2000ec,Gubser:2000mm}.  As pointed out in \cite{Buchel:2005nt}, $C>0$ is equivalent to $c_s^2>0$, which makes the CSC seem inevitable, at least in the absence of conserved charges and in the presence of some kind of holographic dual.  The argument of \cite{Buchel:2005nt} can probably be extended to cover the general case by considering the dispersion relations for all the hydrodynamical modes, including those arising from the dual conserved currents.  However, the CSC remains a conjecture, and there appears to be room for violations: see for example \cite{Friess:2005zp,Marolf:2004fya,Bostock:2004mg}.

The CSC relates only to the existence of a linearized instability around a static or stationary horizon.  Considerable work has been devoted to the question of what the endpoint of the evolution of the Gregory-Laflamme instability might be: see for example \cite{Horowitz:2001cz,Gubser:2001ac,Harmark:2002tr,Wiseman:2002zc,Kol:2002xz,Kol:2004ww}.  When there are thermodynamically stable horizons both with larger and smaller entropy, it seems to us likely that the endpoint of the evolution is a mixed phase with uniformly small curvatures outside the horizon, which remains unbroken.  A mixed phase is a configuration where high entropy and low entropy regions with the same temperature are separated by domain walls.  Typical solutions may not be static, but may instead evolve slowly according toward larger domains according to an effective theory with domain walls whose width is eventually negligible compared to the size of the domains.  Mixed phases were previously suggested in connection with the Gregory-Laflamme instability in \cite{Cvetic:1999rb}.

\begin{figure}
  \centerline{\includegraphics[width=7in]{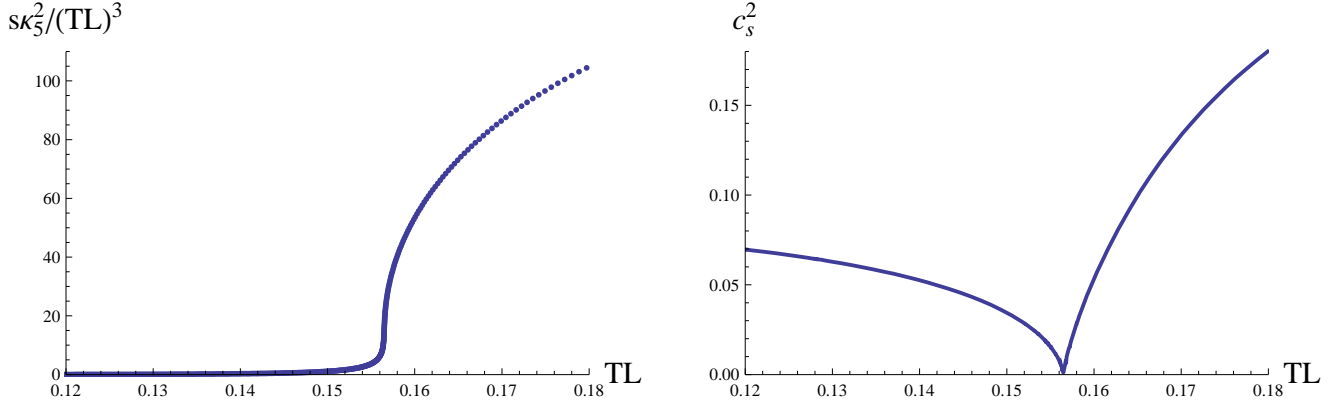}}
  \caption{The equation of state and the speed of sound for $V(\phi) = -{12 \over L^2} \cosh {\phi \over \sqrt{2}} + {1.942 \over L^2} \phi^2$.  The point where $c_s^2 = 0$ is a second order phase transition.  If one considers instead $V(\phi) = -{12 \over L^2} \cosh {\phi \over \sqrt{2}} + b\phi^2$ for $b > 1.942$, the transition becomes first order, while if $b < 1.942$, it is a cross-over.}\label{ex4}
 \end{figure}

Finally, it is possible to arrange second order behavior by tuning the potential $V(\phi)$ so that $c_s^2$ goes to $0$ at some value of $\phi_H$ but never becomes negative: see figure~\ref{ex4}.
There is a corresponding critical temperature, and near it the equation of state typically takes the form
 \eqn{sCritical}{
  s \approx s_0 + s_{1/3} t^{1/3} \qquad\hbox{where}\qquad
   t = {T-T_c \over T_c} \,.
 }
The specific heat diverges as $C \sim t^{-2/3}$, and consequently the speed of sound behaves as $|t|^{-1/3}$.

\section{Discussion}
\label{DISCUSSION}

Since the inception of the anti-de Sitter / conformal field theory correspondence \cite{Maldacena:1997re,Gubser:1998bc,Witten:1998qj}, it has been hoped that it would help solve quantum chromodynamics (QCD).  This hope was articulated most clearly in the early literature in \cite{Witten:1998zw}.  Subsequently, a large and somewhat heterogeneous literature has grown up around the idea of ``AdS/QCD.''  Points of entry into this literature include \cite{Csaki:1998qr,Polchinski:2001tt,Sakai:2004cn,Erlich:2005qh}.

The first thermodynamic question one might ask of a putative dual to QCD is whether the equation of state is right.  We have shown that the equation of state can be built into the construction by choosing an appropriate potential $V(\phi)$ for a scalar field that describes the breaking of conformal invariance.  Indeed, within certain limitations, any equation of state $s = s(T)$ can be translated into a choice of $V(\phi)$, and vice versa.  The limitations include that we use the supergravity approximation.  This immediately points to a weakness of our approach: the shear viscosity will always satisfy $\eta/s = 1/4\pi$, regardless of temperature \cite{Policastro:2001yc,Herzog:2002fn,Buchel:2003tz,Kovtun:2004de}.  Low shear viscosity is in conflict with expectations for the low-temperature phase of QCD, where the mean free path becomes large.  Another reason to be suspicious of any attempt to describe the low-temperature phase using a black hole horizon is that at large $N$, entropy of a horizon scales as $N^2$, whereas the number of degrees of freedom in the confined phase of an $SU(N)$ gauge theory scales as $N^0$.  A black hole description may be approximately valid above $T_c$, and its validity may fail only gradually as one passes through the cross-over.  But sufficiently far below the transition, the paradigm of weakly interacting hadrons should take over, and that is not part of our construction.  One might imagine improvements on our construction, where, for example, higher curvature corrections significantly increase $\eta/s$, especially around or below the transition temperature.  Eventually---perhaps when curvatures near the horizon become sufficiently large compared to the string scale---there could be a cross-over to a gas of strings in a curved spacetime.

Our methods for constructing black holes are more general than the particular problem of mimicking the equation of state of QCD.  Smooth cross-overs, second-order transitions, first-order transitions, and perhaps even mixed phases may all be accommodated within the framework we have proposed.  Our nonlinear master equation approach is special to the case of a single scalar, and it takes advantage of a weak form of integrability of the underlying equations.  However, it is straightforward in principle to work with multiple scalars as well as with gauge fields: in this connection see for example \cite{Gubser:2005ih,Gubser:2008px}.  It seems likely that black holes in suitably designed theories exhibit a remarkable diversity of phase transitions.

\section*{Acknowledgments}

We thank G.~Michalogiorgakis and S.~Pufu for collaboration on early stages of this project, and F.~Rocha for useful discussions.  This work was supported in part by the Department of Energy under Grant No.\ DE-FG02-91ER40671 and by the NSF under award number PHY-0652782.  The work of A.N.~was also supported by the Office of Naval Research via an NDSEG Fellowship.

\bibliographystyle{ssg}
\bibliography{sound}
\end{document}